# Focusing polychromatic light through strongly scattering media

**Hari P. Paudel, Chris Stockbridge, Jerome Mertz, Thomas Bifano**[*]

*Boston University Photonics Center, Boston, MA, USA*
*[*]Corresponding author: tgb@bu.edu*

**Abstract:** We demonstrate feedback-optimized focusing of spatially coherent polychromatic light after transmission through strongly scattering media, and describe the relationship between optimized focus intensity and initial far-field speckle contrast. Optimization is performed using a MEMS spatial light modulator with camera-based or spectrometer-based feedback. We observe that the spectral bandwidth of the optimized focus depends on characteristics of the feedback signal. We interpret this dependence as a modification in the number of spectral modes transmitted by the sample, and introduce a simple model for polychromatic focus enhancement that is corroborated by experiment with calibrated samples.

## References

1. A. Ishimaru, *Wave Propagation and Scattering in Random Media*, IEEE Press Series, (1999).
2. I. M. Vellekoop and A. P. Mosk, "Focusing coherent light through opaque strongly scattering media," Opt. Lett. 32(16), 2309-2311 (2007).
3. S. M. Popoff, G. Lerosey, M. Fink, A. C. Boccara, S. Gigan, "Controlling light though optical disordered media: transmission matrix approach", New J. Phys. 13, 123021 (2011).
4. A. P. Mosk, A. Lagendijk, G. Lerosey and M. Fink, "Controlling waves in space and time for imaging and focusing in complex media," Nat. Photonics, 6 (2012).
5. I. M. Vellekoop, A. P. Mosk, "Phase control algorithms for focusing light through turbid media", Opt. Commun. 281, 3071-3080 (2008).
6. I. M. Vellekoop, A. Lagendijk and A. P. Mosk, "Exploiting disorder for perfect focusing," Nature Photon. 4, 320-322 (2010).
7. S. M. Popoff, G. Lerosey, R. Carminati, M. Fink, A. C.Boccara, and S. Gigan, "Measuring the Transmission Matrix in Optics: An Approach to the Study and Control of Light Propagation in Disordered Media," Phys. Rev. Lett. 104, 100601 (2010).
8. C. Stockbridge, Y. Lu, J. Moore, S. Hoffman, R. Paxman, K. Toussaint, and T. Bifano, "Focusing through dynamic scattering media," Opt. Express, 20(14), 15086-15092 (2012).
9. S. Tripathi, R. Paxman, T. Bifano, K. C. Toussaint, "Vector transmission matrix for the polarization behavior of light propagation in highly scattering media," Opt. Express 20, 16067-16076 (2012).
10. J. Aulbach, B. Gjonaj, P. M. Johnson, A. P. Mosk, and A. Lagendijk, "Control of Light Transmission through Opaque Scattering Media in Space and Time," Phys. Rev. Lett. 106, 103901 (2011).
11. O. Katz, E. Small, Y. Bromberg and Y. Silberberg, "Focusing and compression of ultrashort pulses through scattering media," Nature Photonics 5, 372–377 (2011).
12. D. J. McCabe, A. Tajalli, D. R. Austin, P. Bondareff, I. A. Walmsley, S. Gigan, B. Chatel, "Spatio-temporal focusing of an ultrashort pulse through a multiply scattering medium", Nat. Commun. 2, 447 (2011).
13. D. B. Conkey, R. Piestun, "Color image projection though a strongly scattering wall", Opt. Express, 20, 27312-27318 (2012).
14. E. Small, O. Katz, Y. Guan and Y. Silberberg, "Spectral control of broadband light through random media by wavefront shaping," Opt. Lett. 37(16) (2012).
15. F. Lemoult, M. Fink and G. Lerosey, "A polychromatic approach to far-field superlensing at visible wavelengths," Nat. Comm. 3, 889 (2012).
16. J. W. Goodman, *Speckle Phenomena in Optics*, Roberts & Company, Englewood, CO (2007).
17. D. J. Thouless, 'Maximum Metallic Resistance in Thin Wires," Phys. Rev. Lett. 39, 1167 (1977).
18. S. T. Hong, I. Sreenivasiah and A. Ishimaru, "Plane wave pulse propagation though random media", IEEE Trans. Antennas & Propagation, AP-25, 822-828 (1977).

19. R. Pnini and B. Shapiro, "Fluctuations in transmission of waves through disordered slabs", Phys. Rev. B, 39, 6986-6993 (1989).
20. A. Z. Genack and J. M. Drake, "Relationship between optical intensity, fluctuations and pulse propagation in random media", Europhys. Lett. 11, 331-336 (1990).
21. J. F. de Boer, M. P. van Albada, A. Lagendijk, "Transmission and intensity correlations in wave propagation through random media", Phys. Rev. B, 45, 658-666 (1992).
22. A. M. Weiner, "Femtosecond pulse shaping using spatial light modulators", Rev. Sci. Instrum. 71, 1929 (2000).
23. S. Prahl, Mie scattering calculator (http://omlc.ogi.edu/calc/mie_calc.html).
24. C. A. Thompson, K. J. Webb, and A. M. Weiner, "Diffusive media characterization with laser speckle", Appl. Opt. 36, 3726-3734 (1997).
25. N. Curry, P. Bondareff, M. Leclercq, N. F. van Hulst, R. Sapienza, S. Gigan and S. Gresillon, "Direct determination of Diffusion Properties of Random Media from Speckle Contrast," Opt. Lett. 36 (17) (2011).
26. F. van Beijnum, E. G. van Putten, Ad Lagendijk, and A. P. Mosk, "Frequency bandwidth of light focused through turbid media," Opt. Lett. 36, 373-375 (2011).
27. J. Aulbach, A. Bertagne, M. Fink, M. Tanter and A. Tourin, "Optimal spatiotemporal focusing through complex scattering media," Phy. Rev. E 85, 016605 (2012).
28. M. A. Webster, K. J. Webb, A. M. Weiner J. Xu, H. Cao, "Temporal response of a random medium from speckle intensity frequency correlations", J. Opt. Soc. Am. 20, 2057-2070 (2003).
29. F. Helmchen, W. Denk, "Deep tissue two-photon microscopy," Nat. Method 2(12) (2005).
30. M. J. Booth, "Adaptive optics in microscopy," Philos. Transact. A Math. Phys. Eng. Sci. 365(1861), 2829–2843 (2007).
31. J. Tang, J. N. Germain, and M. Cui, "Superpenetration optical microscopy by iterative multiphoton adaptive compensation technique," PNAS, 109(22) 8434-8439 (2012).
32. J. Aulbach, B. Gjonaj, P. Johnson, A. Lagendijk, "Spatiotemporal focusing in opaque scattering media by wavefront shaping with nonlinear feedback", Opt. Express 20, 29237 (2012).
33. D. Débarre, E. J. Botcherby, T. Watanabe, S. Srinivas, M. J. Booth, and T. Wilson, "Image-based adaptive optics for two-photon microscopy," Opt. Lett. 34(16), 2495–2497 (2009).
34. N. Ji, D. E Milkie and E. Betzig, "Adaptive optics via pupil segmentation for high-resolution imaging in biological tissues," Nat. Methods 7, 141 - 147 (2010).
35. X. Xu, H. Liu and L. V. Wang, "Time-reversed ultrasonically encoded optical focusing into scattering media," Nat. Photonics 5, 154–157 (2011).

## 1. Introduction

The problem of focusing light through thick, scattering media has been longstanding, particularly in the biology community where transport scattering lengths are typically in the range of millimeters. Beyond this range, light is considered to be in a diffusion regime [1], meaning that essentially no ballistic light is available to form a focus. Nevertheless, it was demonstrated recently that that by controlling the phase front of light with a large number of spatial degrees of freedom (e.g. segments of a spatial light modulator (SLM)), the effects of scattering can be partially compensated and the ability to form a bright focal spot can be partially restored [2]. This groundbreaking demonstration has launched a wave of research interest [see review articles 3, 4].

Most studies on the focusing of light through strongly scattering media make use of monochromatic light [2, 5-9]. With no focus optimization, a spatially coherent monochromatic light beam entering a medium emerges with a randomized phase front, such that in the far field the light produces a fully developed speckle pattern of unit contrast. When focus optimization is applied to increase the intensity at a single spot in this pattern, an intensity enhancement is achieved roughly equal to the number of pixels $N$ in the SLM (provided that these pixels constitute independent degrees of freedom). Phenomenologically, it has been found that when focus optimization is applied to not one but $M$ distinct spots simultaneously, the intensity enhancement at each point is given by roughly $N/M$ [2]. This follows from the argument that the SLM pattern that optimizes one spot is uncorrelated with the SLM pattern that optimizes another. As a result, only $N/M$ degrees of freedom can be dedicated to the optimization of each spot, and the enhancement per spot is commensurately reduced.

More recently, focus optimization experiments have begun to explore the use of non-monochromatic light sources [10-15] (referred to here as polychromatic). When spatially coherent polychromatic light is transmitted through a scattering medium, it too generates a speckle-like pattern in the far field, but with reduced contrast compared to a monochromatic beam. Indeed, the speckle-like pattern can be thought of as the incoherent superposition of multiple uncorrelated speckle patterns produced by the different frequency components encompassed by the light spectrum [16]. The number of frequency components, or spectral modes, contributing to this superposition depends on the spectral bandwidths of both the illumination and the sample (the sample bandwidth is defined here as the inverse Thouless time [17-21], roughly given by $D/L^2$, where $D$ is the diffusion constant of light in the sample and $L$ is the sample thickness).

Focus optimization of polychromatic light thus corresponds to the local intensity optimization of a superposition of multiple independent spectral modes, each effectively monochromatic, using a single SLM. At first glance, this problem appears equivalent to the problem of simultaneous focus optimization of multiple independent focal spots as described above. One might therefore expect an intensity enhancement factor for polychromatic focus optimization to be given by $N/M$, where $M$ is the number of spectral modes encompassed by the polychromatic beam [14,15]. In practice we have found this relation to be true, but only provided one makes allowances for an effective broadening of the sample bandwidth dependent on the type of feedback used for focus optimization. The purpose of this work is to provide evidence for this broadening and discuss some of its ramifications.

## 2. Experimental Procedure

Our experiments consisted of an optical apparatus in which spatially coherent illumination of variable spectral bandwidth was transmitted through a variety of thicknesses of strongly scattering samples. The spatial phase of the illumination beam was controlled using a MEMS SLM. Feedback for focus optimization was mediated either by a camera in the far field (detailed in Section 3) or by a spectrometer in the far field (detailed in Section 4).

A schematic of our apparatus is shown in Fig.1. The broadband source is a 5mW continuous-wave superluminescent diode (Superlum SLD-33-HP) coupled into single-mode fiber. The bandwidth of this SLD is 14.4THz with central frequency 379THz. The diverging beam at the SLD fiber output is collimated by lens L1 ($f = 25$mm). This illumination beam is transmitted through a polarizing beam splitter and a quarter wave plate. The spectral bandwidth of the beam is adjusted using a standard double-pass configuration [22] with a diffraction grating (1200 lines/mm), lens L2 ($f = 200$mm) and a variable-width slit. A beam expander comprised of two lenses L3 ($f = 25$mm) and L4 ($f = 100$mm) magnifies the beam so that it roughly fills the aperture of a high-speed micro-electromechanical SLM (Boston Micromachines Corporation Kilo-SLM). The area of the SLM covered by the illumination beam encompasses approximately 900 reflective square segments (pixels), each with surface area 0.09 mm$^2$. The beam is then focused onto the front surface of a strongly scattering sample using two lenses L5 ($f = 62$mm) and L6 ($f = -25$ mm) and a 10× 0.3NA microscope objective. Light transmitted through the sample is collected by a 20× 0.4NA microscope objective with tube lens L7 ($f = 125$mm), to produce a linearly polarized (P) speckle pattern at the sensor of a CMOS camera (μEye USB 2LE). The speckle grain size is controlled with an adjustable aperture located between the polarizer and the tube lens. Four calibrated scattering samples of thicknesses $L$ equal to 5.2$l$*, 2.1$l$*, 1.1$l$* and 0.8$l$* were used in the experiments, where $l$* is the transport mean free path in the sample. The first two samples were made from a suspension of 1μm diameter polystyrene beads; the last two samples were made from a suspension of 6.5μm diameter borosilicate beads. In all cases these were embedded in a silicone matrix (Dow Sylgard 184), and the associated values of *l** were calculated from Mie theory [23].

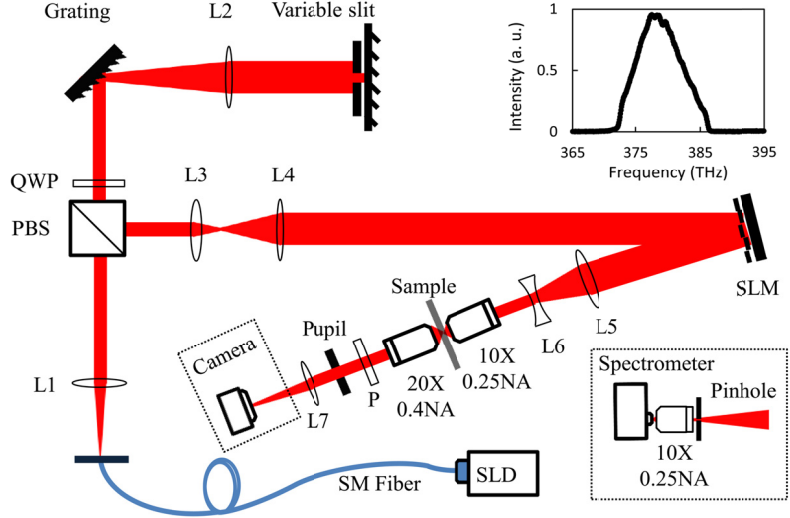

Fig. 1. Schematic of the experimental setup. The spectrum (top right inset) of a fiber-delivered broadband SLD is adjusted using a grating and variable-width slit before being reflected from a MEMS spatial light modulator (SLM) and focused onto the surface of a strongly scattering sample. Transmitted light produces a speckle pattern that is sensed by either a camera or a spectrometer (bottom right inset). Focus optimization is based on a coordinate descent optimization algorithm [8]. SM = single mode; L1-L7 = lenses; PBS = polarizing beamsplitter; QWP = quarter-wave plate; P = polarizer.

The illumination spectral bandwidth for different slit widths was measured with a spectrometer inserted in front of the SLM (not shown in Fig. 1), and calculated from

$$\Delta \nu_l = \frac{\left(\int S_0(\nu) d\nu\right)^2}{\int S_0^2(\nu) d\nu}, \quad (1)$$

where $S_0(\nu)$ was the measured spectral profile as a function of frequency $\nu$. The advantage of this definition of bandwidth is that it is less sensitive to the exact shape of the spectral profile than, say, a definition based in full-width at half maximum. The bandwidths used in experiment ranged from 11.8THz at the maximum (7mm) slit width to 2.2THz at the minimum (1mm) slit width. Experiments were also conducted with a near monochromatic source: a 4.5mW laser diode (Thorlabs CSP192) with 0.35THz bandwidth and 386THz central frequency (not shown in Fig. 1). In those experiments, the diode laser beam was spatially filtered using a 4μm pinhole and coupled into a single mode fiber using a 5× 0.1NA microscope objective.

Experiments consisted of initial measurements of the speckle contrast at the camera sensor for a variety of combinations of illumination bandwidths and scattering samples, followed by the implementation of an optimization loop aimed at increasing the intensity of a localized spot in the transmitted light pattern. The spot size was chosen to be slightly smaller than the characteristic size of a speckle grain. The optimization algorithm sequentially maximizes the spot intensity as a function of phase coefficients for each of 1024 orthogonal Hadamard modes. Details of this optimization algorithms were reported previously [8]. The intensity enhancement achieved with spot optimization is defined here by $E=I_{opt}/I_{avg}$, where $I_{opt}$ is the maximum spot intensity attained with feedback optimization, and $I_{avg}$ is the local average intensity without feedback optimization [2].

## 3. Camera-based feedback

To begin, we performed focus optimization with camera mediated feedback and variable bandwidth illumination, as shown in Fig. 1. To determine the number $M$ of spectral modes transmitted through the sample, we measured the spatial contrast of the speckle pattern at the camera sensor before optimization feedback. Contrast is defined here by $C=\sigma/I_{avg}$, where $\sigma$ and $I_{avg}$ are the intensity standard deviation and average, respectively, measured over a region of the camera sensor where the intensity pattern statistics appeared spatially homogeneous [16]. For polychromatic light, the measured contrast is given by $C_0/\sqrt{M}$, where $C_0$ is the measured contrast produced by monochromatic light. In theory, $C_0$ should be equal to 1 for a fully developed monochromatic speckle pattern, however in our case it was measured with the laser-diode source to be closer to 0.87, owing to the spatial filtering caused by the non-zero pixel sizes of the camera sensor. Correcting for this filtering, we have then for polychromatic light

$$C = \frac{1}{\sqrt{M}}. \qquad (2)$$

We measured contrast values $C$ as a function of $\Delta\nu_l$ for various samples. The results are shown in Fig. 2. While the relation between $C$ and $M$ is straightforward for polychromatic light, the relation between $C$ and $\Delta\nu_l$ is not. In the latter case, $C$ depends on the specific spectral profiles not only of the illumination beam but also of the spectral modes encompassed by this beam, which in turn depends on the sample itself [16, 24, 25]. For our purposes we make use of a very simplified model based on our experimental results. These are found to fit the approximation

$$C = \sqrt{\frac{\Delta\nu_s}{\Delta\nu_l + \Delta\nu_s}} \qquad (3)$$

where $\Delta\nu_s$ corresponds to the bandwidth of the spectral modes transmitted by the samples (i.e. the inverse Thouless time), in accord with our independent assessments of $D$ and $L$. In addition to providing an excellent fit to data, this approximation trends toward the limits $M\rightarrow 1$ for $\Delta\nu_l \rightarrow 0$ (monochromatic illumination) and $M\rightarrow \Delta\nu_l/\Delta\nu_s$ for $\Delta\nu_l >> \Delta\nu_s$ (broadband illumination or very thick sample), as expected.

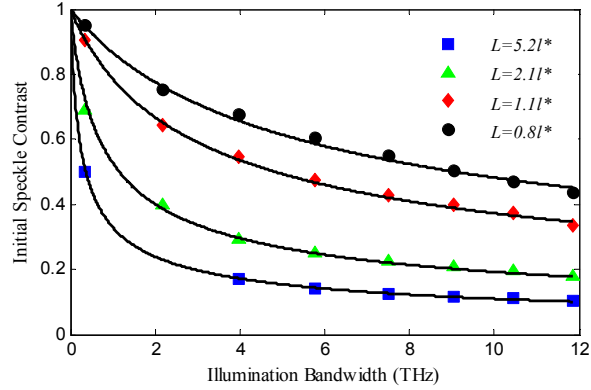

Fig. 2. Measured contrast ($C$) as a function of illumination spectral bandwidth ($\Delta\nu_l$) for four samples of thicknesses $L = 5.2l^*$, $2.1l^*$, $1.1l^*$ and $0.8l^*$. The plots are fits to the experimental data based the model Eq. (3).

Having determined $M$ from our contrast measurements, we now evaluate the effect of $M$ on the intensity enhancement $E$ that can be achieved with focus optimization of polychromatic light. As argued in the introduction, one might expect $E = E_0/M = E_0/C^2$, where $E_0$ is the maximum enhancement that can be achieved with monochromatic illumination (given by $E_0 = \pi/4\,N$, where $N \sim 900$ is the number of pixels used in the SLM [2]). The results are shown in Fig. 3.

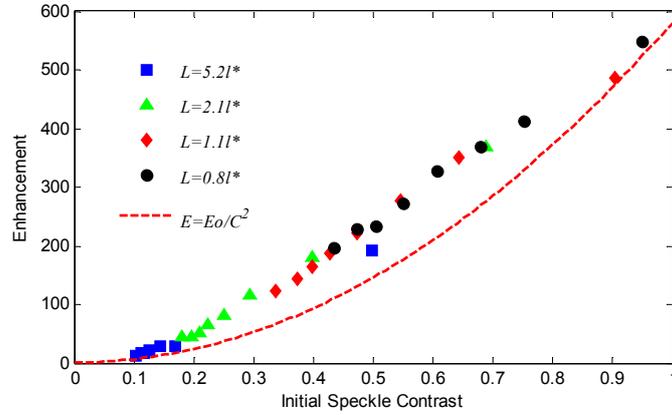

Fig. 3. Optimized focus intensity enhancement ($E$) versus measured initial speckle contrast ($C$) for combinations of samples and illumination bandwidths. The red dashed trace shows $E=E_0/C^2$, for reference.

Two features in Fig 3 are of note. The first is that the enhancement factor $E$ does indeed appear to obey a universal curve dependent exclusively on the initial speckle contrast $C$ and independent of sample specifics. However, the second note is that this curve is not the expected curve (dashed trace). Manifestly, the measured enhancement achieved with polychromatic light is somewhat greater than first expected. The purpose of the remainder of this paper is to investigate this discrepancy.

**4. Spectrometer-based feedback**

To better understand the nature of the focus optimization implemented above, we isolated a single speckle grain, or rather the area corresponding to a single speckle grain, with a pinhole (75μm diameter) placed at the location of the camera sensor. We further replaced the camera sensor with a microscope objective (10× 0.25NA) and spectrometer (Thorlabs CCS175), as shown in the inset of Fig. 1. In this manner, we could monitor the spectrum $S(\nu)$ of a single transmitted spatial mode (spatial speckle grain). The integral of this spectrum over all frequencies corresponds to a focal spot intensity as would be measured by a camera. Focus optimization based on maximizing this integrated spectrum is thus equivalent to focus optimization with a camera sensor, and we expect identical results as shown in Fig. 3. Spectrometer-based feedback was utilized previously [12,14].

An example of a spot spectrum prior to feedback optimization and with full illumination bandwidth is shown in Fig. 4 (black trace). As expected, this spectrum is granular and exhibits what are commonly referred to as spectral speckles. The characteristic bandwidth of these spectral speckles corresponds to the range over which the transmitted light is effectively monochromatic. More specifically, this characteristic bandwidth is equal to $\Delta\nu_s$. For the example shown in Fig. 4, $\Delta\nu_s$ is expected to be 1.6THz as inferred from the measurement of spatial speckle contrast (see Fig. 2). This is consistent with the characteristic spectral speckle bandwidth inferred from our contrast measurement and roughly observed in Fig. 4. For this example, we thus have $M \approx 8$.

An advantage of spectrometer-based feedback is that it provides additional degrees of freedom for focus optimization. For example, instead of maximizing the integral of the full spectrum $S(\nu)$, we can maximize the integral of only a portion of this spectrum. That is, we can arbitrarily control the signal bandwidth used for focus optimization. Fig. 4 shows an example of this. We began by performing focus optimization with a narrow signal bandwidth smaller than the characteristic spectral speckle bandwidth $\Delta\nu_s$. That is, though the incident light was polychromatic, only a monochromatic portion of this light (i.e. a single spectral mode) was used for feedback. Focus optimization in this case led to a new focus spectrum $S_{opt}(\nu)$ (red trace in Fig. 4), and to a significantly enhanced focus intensity. The observed final bandwidth of $S_{opt}(\nu)$ was found to somewhat larger than $\Delta\nu_s$ by a factor of approximately 1.6. This is consistent with a previous report using monochromatic light [26], where the spectrum $S_{opt}(\nu)$ was found to coincide with the correlation of a spectral speckle, which has a larger bandwidth than spectral speckle itself ($\Delta\nu_s$).

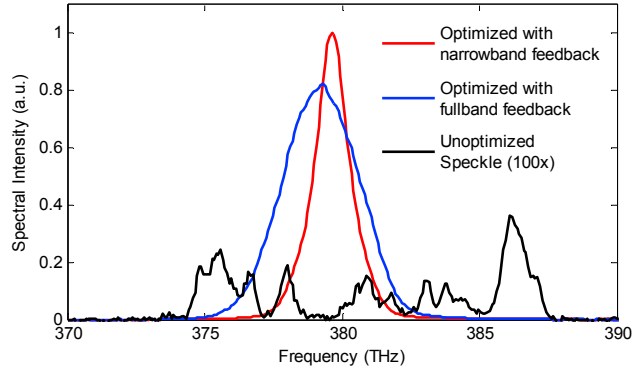

Fig. 4. Spectral profiles of a single speckle grain before (black) and after focus optimization using first a narrowband feedback signal (red) and then a full bandwidth feedback signal (blue). The feedback signal bandwidths were 0.4THz and 11.8THz, respectively. The resulting optimized speckle bandwidths were 2.6THz and 5.3THz, respectively. The sample thickness was $L = 1.1l^*$. From the black curve, we observe that a single un-optimized spatial speckle grain contains several spectral speckle modes. From Eq. 3, the bandwidth of these modes was calculated to be $\Delta\nu_s$ =1.6THz.

Less expected was the outcome of the second part of our experiment where, starting from the final SLM pattern attained with narrowband signal optimization, we then switched to full bandwidth signal optimization and continued to apply our feedback algorithm. The final result is shown in Fig. 4 (blue trace). The continued signal optimization led to a slight reduction in the peak height of $S_{opt}(\nu)$ accompanied by a significant spectral broadening, the net effect being a further increase in focus intensity (full integral of $S_{opt}(\nu)$). Manifestly, full bandwidth signal optimization, equivalent to camera-based focus optimization as performed in Section 3, led to the enhancement of frequencies over a broader spectral range than expected from $\Delta\nu_s$. This result suggests that full bandwidth signal optimization naturally "prefers" to distribute intensity enhancement over a larger frequency range than encompassed by a single spectral mode. We note that similar results as the blue trace in Fig 4 were obtained when starting the feedback optimization from arbitrary SLM patterns, with the difference that the enhanced focus spectra were centered about arbitrary frequencies (generally clustered near the peak of the SLD spectrum, but sometimes not).

The propensity of full bandwidth signal optimization to distribute intensity enhancement over more than one spectral mode was controlled by yet another degree of freedom in our optimization algorithm. Specifically, a further generalization of the optimization metric used for spectrometer-based feedback is given by

$$J = \int_B S^\alpha(\nu) d\nu \tag{4}$$

where $J$ is the scalar metric that is maximized, $B$ defines the range over which the focus spectrum is integrated (i.e. the signal bandwidth), and $\alpha$ is a new parameter we have introduced. So far, we have considered optimization metrics with various bandwidths $B$, but we have only considered the linear case $\alpha = 1$. As a reminder, standard camera-based focus optimization as performed in Section 3 corresponds to full $B$ and $\alpha = 1$.

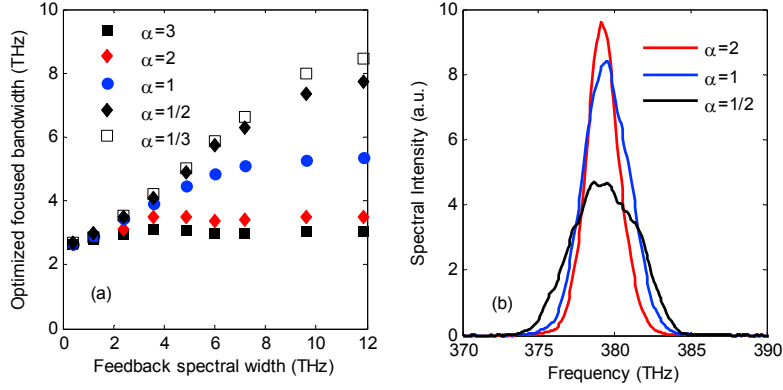

Fig. 5. (a) Spectral bandwidth of the optimized focus as a function of increasing feedback signal bandwidth $B$, for different values of the nonlinearity parameter $\alpha$. (b) Spectral profile of representative optimized foci for full bandwidth feedback and different values of $\alpha$.

Fig. 5 summarizes the results obtained when using signal feedback with various values of $\alpha$. Specifically, Fig. 5a shows the bandwidth of the resultant focus spectrum when performing signal optimization with progressively larger signal bandwidths. As can be seen, for narrowband signal feedback, the optimized focus spectrum is equal to $\Delta\nu_s$, independent of $\alpha$. As the signal bandwidths become progressively larger, the optimized focus bandwidths also become larger but then plateau at a level governed by $\alpha$. Sub-linear values of $\alpha$ lead to large increases in the optimized focus bandwidth (factors of 3 or greater), whereas supra-linear values of $\alpha$ hardly lead to increases at all. An explanation for the results observed in Fig. 5a is that sub-linear values of have the effect of "flattening" the feedback signal spectrum and thus facilitating the distribution of intensity enhancement over a larger spectral range. Supra-linear values of, in turn, have the opposite effect. Fig. 5b shows representative optimized focus spectra for different values of $\alpha$. We note that optimization with $\alpha = 1/2$ is similar to what was previously referred to as "amplitude optimization" [5], though here the amplitude is spectral rather than spatial.

We recall that $\Delta\nu_s$ corresponds to the frequency range over which the light transmitted through the sample can be regarded as monochromatic. The results above suggest that this frequency range is not a fixed value for a given sample (dependent only on $D$ and $L$), but rather that it can be manipulated by the optimization feedback itself, for example by adjusting the parameters $B$ or $\alpha$. That is, upon optimization feedback, $\Delta\nu_s$ becomes modified. We denote this modified bandwidth as $\xi\Delta\nu_s$, where $\xi$ is equal to one for narrowband signal feedback ($B$ small), but is in general greater than one for broadband signal feedback ($B$ large). In the case of broadband signal feedback with $\alpha = 1$ (i.e. standard focus optimization), we observe from Fig. 5a that $\Delta\nu_s$ is approximately doubled upon optimization, meaning that $\xi$ is approximately equal to 2.

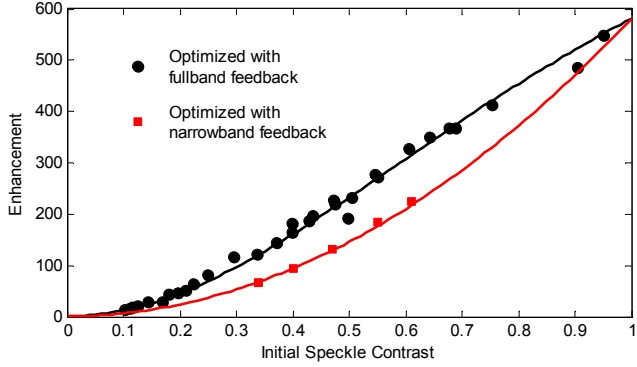

Fig. 6. Enhancement ($E$) versus contrast ($C$) plots. Filled circles are enhancements obtained using camera-based feedback with $\alpha = 1$ and full $B$ (same as Fig. 2). Filled diamonds are enhancements obtained under the same experimental condition using spectrometer-based feedback with $\alpha = 1$ and small $B$ (less than $\Delta\nu_s$). Corresponding plots are derived from Eq. 5 (i.e. they are not fits).

To evaluate the consequences of an effective broadening of $\Delta\nu_s$ on focus enhancement, we return our simple model given by Eq. 3. A broadening of $\Delta\nu_s$ leads to a modification of the effective number of spectral modes contained in the transmitted polychromatic focal spot. Specifically, we find then

$$M_{eff} = \frac{1}{\xi}(M-1)+1 = \frac{1}{\xi}(C^{-2}-1)+1, \tag{5}$$

where $C$ is the spatial speckle contrast prior to optimization, as before. For monochromatic illumination, $M_{eff}$ remains equal to 1. However, for polychromatic illumination, the expected focus enhancement now becomes $E = E_0/M_{eff}$. A plot of this expected enhancement is presented alongside our experimental data from Fig 2. The fit now appears accurate (black trace). To further validate our model, we performed focus optimization but with a narrowband feedback signal ($B = 0.4$THz) rather than a full bandwidth feedback signal. In this case $\xi = 1$ and we recover an enhancement given by $E = E_0/M$ (red trace). Again, the model appears to fit our data.

## 5. Conclusion
In conclusion, we have shown that the enhancement factor obtained when applying focus optimization to a polychromatic beam transmitted through a strongly scattering medium depends both on the number of spatial and spectral degrees of freedom in the optimization feedback. It is well known that spatial and temporal degrees of freedom are tightly coupled when performing optimization feedback [10-12, 20], and this result is therefore not surprising. What is less expected is the observation that the number of spectral degrees of freedom seems itself to be a function of the optimization feedback parameters. In particular, we have shown that measured values of focus enhancement can be explained by a phenomenological broadening of $\Delta\nu_s$. Recalling that $\Delta\nu_s$ is related to the inverse Thouless time, or equivalently, to the width of the optical pathlength distribution of random light trajectories through the scattering medium [28], we may speculate that the effective broadening of $\Delta\nu_s$ is accompanied by a concomitant narrowing of the optical pathlength distribution width. This remains to be verified experimentally.

The results presented here are relevant to the problem of focusing polychromatic light in thick tissue, as is encountered in a variety of applications for example involving nonlinear optical interactions [29-35] in which optical scattering imposes practical limits on achievable imaging depths.

**Acknowledgments**

This work was supported by National Academies Keck Futures Initiative (NAKFI) grant number NFAKI-IS10, W.M. Keck Foundation. We are grateful to Boston Micromachines Corporation (BMC) for providing the SLM. Thomas Bifano acknowledges a financial interest in BMC. We thank R. Barankov and J. D. Giese for sample preparation and calibration.